\documentstyle[preprint,eqsecnum,aps]{revtex}

\begin{document}
\draft
\preprint{}

\title{Scaling Phenomena in a Unitary Model of Directed\\ Propagating Waves
 with Applications to One-dimensional\\ Electrons in a Time-varying Potential}

\author{Dinko Cule and Yonathan Shapir}

\address{
 Department of Physics and Astronomy\\University of Rochester\\
 Rochester, NY 14627}

\date{\today}
\maketitle
\begin{abstract}

We study a 2D lattice model of forward-directed waves in which the integrated
intensity for classical waves (or probability for quantum mechanical particles)
is conserved. The model describes the time evolution of 1D quantum particle in
a
time-varying potential and also applies to propagation of electromagnetic waves
in two dimensions within the parabolic approximation. We present a closed form
solution for propagation in a uniform system. Motivated by recent studies of
non-unitary directed models for localized 2D electrons tunneling in a magnetic
field, we then address related theoretical questions of how the interference
pattern between constrained-forward paths in this unitary model is affected
by the addition of phases corresponding to such a magnetic field.
The behavior is found to depend sensitively on the
value of $\Phi/\Phi_{0}$, where $\Phi$ is flux per plaquette and $\Phi_{0}$ is
the unit of flux quantum. For $\Phi/\Phi_{0} = p/q$ we find the amplitude to be
more collimated the larger is the value of $q$. We next consider propagation in
a random forward scattering media. In particular the scaling
properties associated
with the transverse width $x$ of the wave, as function of its distance $t$ from
point source, are addressed. We find the moments of $x$ to scale with $t$ in a
very different way from what is known for either off-lattice unitary or
on-lattice non-unitary systems. The scaling of the moments of the
probability $[P^{n}(x,t)]$ (or intensity) at a point $(x=0,t)$
is found to be consistent with a simple behavior
$[P^{n}(0,t)] \sim t^{-\frac{n}{2}}$. Implications to the behavior of
one-dimensional lattice quantum particles in a dynamically fluctuating
random potential are discussed.
\end{abstract}


\narrowtext

\section{Introduction}\
\label{sec:Introduction}

   Directed propagation of waves in uniform and nonhomogeneous media has many
applications in different branches of physics.
Classically they arise in the contest of electromagnetic wave propagations
(e.g. light propagation in the atmosphere). Quantum mechanically they arise
in investigations of hopping transport in insulators. If the preferred
direction
is identified as time, it also describe the motion of a quantum particle
which obeys the time-dependent Schr\"{o}dinger equation.

  In the classical waves context directed propagation is obtained within the
so-called "parabolic approximation" which is applied in circumstances at which
most of the scattering  is in the forward direction, [1-5].
The equation is "parabolic"
because the second derivative in the direction of propagation (say $z$) is
replaced by a first one, and the Maxwell equation for a field
$E(x,y,z,t) = Re[{\cal E}(x,y,z,t)e^{i(kz-\omega t)}]$
is replaced by

\begin{equation}
\label{i1}
\left(\,i\frac{\partial}{\partial z} + \frac{1}{2k}\,\nabla^{2} -
 k\,\mu(x,y,z,t) \right)\,{\cal E}(x,y,z,t) = 0
\end{equation}

\noindent
where $\mu(x,y,z,t)$ is a nonuniform index of refraction and
$\nabla^{2} = \frac{\partial^{2}}{\partial x^{2}} +
\frac{\partial^{2}}{\partial y^{2}}$.

    In the applications for classical waves, however, in many instances the
absorption of the medium in which the "directed propagation" takes place
is negligible. Therefore the total integrated energy at every latitude $z$
is constant as energy is conserved during the propagation. We denote by
unitary (for obvious reasons) the models in which $|E|^{2}$ (or $|\Psi|^{2}$
where $\Psi$ is the wave function of a quantum particle)
are conserved.
One of the most important issues we address in this work is the drastic
difference in the properties of directed propagation between unitary and
non-unitary models \cite{FGZ,MKS,SKR}.

   To approach systematically the problems associated with directed waves in
ordered and random media it was found very beneficial to study lattice models.
They are often simpler and more amenable to analytic and numeric
calculations than their continuum counterparts. But for these studies to
universally apply to systems without an underlying lattice structure as well,
the question of whether the continuum behavior is recovered from the lattice
on length scales much longer than the lattice spacing, should be carefully
explored \cite{SKR}.

   In this paper we introduce a two-dimensional lattice model
for unitary directed propagation
and investigate its behavior in uniform, nonhomogeneous and random media.
Important differences between the lattice and the continuum description will
be emphasized. Despite these differences, the study of lattice models helps
in the basic understanding of directed propagation in various environments.
Moreover, it provides a rich collection of phenomena with potential
applications in lattice systems.

   Quantum mechanically the model we study is describing the real time
evolution
of an one-dimensional quantum particle (we shall use electron to denote all of
them) in a time-dependent potential $V(x,t)$:

\begin{equation}
\label{i1-1}
\imath \,\hbar\,\frac{\partial\psi}{\partial t} =
-\,\frac{\hbar^{2}}{2m}\,\frac{\partial^{2}\psi}{\partial x^{2}} + V(x,t)\,\psi
\end{equation}

  We shall study here the case of a random potential V(x,t).
While early studies concentrated on the continuum version \cite{OE,JK} a
lattice model was recently extensively investigated by Bouchaud et al.
\cite{BTS}. To conserve probability on a lattice they employed a different
method than ours. From their numerical investigation they reached the
conclusion that the wave function has multifractal properties. The difference
between the discretization methods and the possibility that they may lead to a
different behavior will be discussed in the last section.

  Another interesting extension we study here is the discrete equqtion:
\begin{equation}
\label{i1-2}
\psi(x,t+1) = e^{\imath\frac{\alpha t}{2}}\,\psi(x-1,t) +
              e^{-\imath\frac{\alpha t}{2}}\,\psi(x+1,t)
\end{equation}

\noindent
The main motivation to study this equation is its close formal connection to
another interesting system: That of tunneling electron on a lattice in presence
of a magnetic field. The directed path approximation is invoked for this case
of
strongly localized electrons, precisely because the wave function decays
exponentially. Therefore the "directed paths" approximation cannot be applied
in the unitary case to study the spatial behavior of the wave function
$\psi(x,y)$ in 2D. However, because of the
interesting behavior \cite{SW2,FSW} found in the
non-unitary case with a magnetic field, there is a strong motivation to widen
our theoretical understanding and to investigate what would be the effect of a
"magnetic field" on the unitary model with constrained forward  paths.
The "magnetic field" is added to the model by adding the "time dependent"
phases (as described in Eq.(\ref{i1-2})) as for the tunneling
electrons in the non-unitary model (in which $t$ denotes
the direction of propagation). It should be emphasized again that the
behavior found does not describe the behavior of non-localized electrons in
a real magnetic field.

  The organization of this paper is as follows: In Sec. 2 the model is
introduced and its behavior on a pure (uniform) lattice is explored in Sec. 3.
Next, we address the question what will happen to the
interference in our model if phases, as if induced by a "magnetic field", are
added to the bonds. That will be discussed in Sec 4.
Sec. 5 is devoted to the effects of disorder on the
propagation of directed waves. The last section (6) is devoted to
concluding remarks.

  After this project was in progress we have learned that the same model
has been studied independently by Saul et. al. in ref. \cite{SKR}.
To some extent our study overlaps with
theirs. In these cases our results also agree with theirs and we therefore
limit their discussions to the necessary minimum in order to make our paper
self-contained. We mostly expand on new directions and extensions beyond
their study.
\section{The Model}
\label{sec:Model}

We shall study the propagation of directed waves on a two dimensional
lattice where the direction
of propagation is identified as the "time" axis. Such a lattice is illustrated
in
Fig. 1. An incoming wave may propagate only along lattice bonds. Each site
represents a scatterer described by some scattering $S$-matrix. All scattering
matrices are elements of the $U(2)$ group (there are two scattering
channels). The wave function $\Psi(t)$ of propagating wave is defined by a set
of $2N$ amplitudes $\psi_{n}(t),\;(n=1,2,\ldots\;,2N;\;N\geq t)$ taking
values on the lattice links. "Time" $t$ corresponds to the number of steps
in the $t$-direction.

  At each site the $S$-matrix transforms two incoming amplitudes into two
outgoing amplitudes. For example, amplitudes $\psi_{n}(t+1)$ and
$\psi_{n+1}(t+1)$, (see Fig. 1), are related to $\psi_{n}(t)$ and
$\psi_{n+1} (t)$ by a matrix multiplication:

\begin{equation}
\label{mo1}
\left( \begin{array}{c} \psi_{n+1}(t+1)\\ \psi_{n}(t+1) \end{array}\right) =
\left( \begin{array}{cc} S_{11}(x,t)&S_{12}(x,t)\\S_{21}(x,t)&S_{22}(x,t)
\end{array}\right)\left( \begin{array}{c}\psi_{n+1}(t)\\ \psi_{n}(t)
\end{array}
\right)\,,
\end{equation}

\noindent
where $x$ is transversal position of scatterer, between
$n^{th}$ and $(n+1)^{th}$ row.

    The $S$-matrix elements are closely related to the
potential $V(\vec x,t)$  in eq. (\ref{i1-1}). One way
to discretize eq. (\ref{i1-1}) is to formulate the problem in terms
of scattering matrices assigned only to the discrete set of space-time
coordinates. The main advantage of this discretization procedure is that
unitarity is manifestly preserved which implies  that the norm
of the wave function is also preserved (there is no dissipation).
A shortcoming of the discretization is the existence of a "light cone" which
precludes any spread faster than  $\langle x^{2} \rangle \sim t^{2}$.
For example, $\langle x^{2} \rangle \sim t^{3}$ found in the continuum,
\cite{OE,JK}, cannot be achieved in this model unless it is generalized to
allow "long range" scatterings (not restricted to nearest neighbors).
\section{Directed waves on  a pure lattice}
\label{sec:pure}

         In this section we  briefly discuss the wave propagation
in the absence of both disorder and  an external field.
The main goal of this study is to describe analytical methods
which will be used in the next sections.

All sites on the lattice will have the same $S$ matrix.
As mentioned earlier, it must be a unitary matrix which we choose
to have all real elements

\begin{equation}
\label{p1}
 S = \frac{1}{\sqrt{2}} \left( \begin{array}{rr} 1 & 1 \\ -1 & 1 \end{array}
                         \right).
\end{equation}
The transfer matrix $T$ relates wave functions  at $t$ and $t-1$:

\begin{equation}
\label{p2}
\Psi(t) = T\,\Psi(t-1)\:,
\end{equation}

\noindent
where $T$ is $2N\times2N$ matrix given by

\begin{equation}
\label{p3}
T = \frac{1}{2}\left( \begin{array}{rrrrrrrr}
       \ddots &    &    &    &    &    &    & \\
              & -1 & -1 & -1 &  1 &    &    & \\
              &  1 &  1 & -1 &  1 &    &    & \\
              &    &    & -1 & -1 & -1 &  1 & \\
              &    &    &  1 &  1 & -1 &  1 & \\
              &    &    &    &    &    &    & \ddots \end{array} \right).
\end{equation}

\noindent
The $T$ matrix
in eq. (\ref{p3}) takes into account scatterings on two
consecutive columns of
sites, (see Fig. 1). Transfer matrices for each individual
column are  actually different. Combining them  together
gives the matrix (\ref{p3}) which then
describes  the transfer over two columns.
Diagonalizing $T$ by a unitary transformation  $T = U^{-1}\,D\,U$,
we have
\begin{equation}
\label{p4}
\Psi(t) = U\,D^{t}\,U^{-1}\:\Psi(t=0).
\end{equation}

Once the wave function
$\Psi(t=0)$ at $t=0$ is known one can obtain the wave function at
arbitrary $t$  from eq. (\ref{p4}).
For example, one can choose as
an initial condition  a wave function whose components on the first links
following the origin have values $1/\sqrt{2}$. This
corresponds to an incident wave  which at $t=0$ which
comes  to the scatterer at  position $x = 0$
from below (see Fig. 1)
and scatters  into upper or lower link
with equal probability 1/2, see (\ref{p1}). Therefore, we have

 \begin{equation}
\label{p5}
\psi_{n}(t=0) = \delta_{n,N}\:,
\end{equation}

\noindent
where $\psi_{n}(t=0)$ is the $n^{\rm th}$ component
of $\Psi(t=0)$, as is shown in Fig. 1.

Note that scattering matrix given by eq. (\ref{p1}) is not symmetric.
This asymmetry of $S$ will produce
the asymmetry among the components of wave function at some  later
$t$. One could obtain a
symmetric expression for the wave function  by
an appropriate superposition of
another possible initial condition, i.e. the
incident wave comes  to the first scatterer from above.
In the presence of disorder
the symmetry will be restored.
We also checked  numerically that  the
choice of initial condition does not have any effect on the scaling exponents.

In order to diagonalize $T$ we impose periodic boundary conditions
and choose for definiteness $N$ to be even. The eigenvalues of $T$
are

\begin{equation}
\label{p6}
 \lambda_{\pm} = -2\sin^{2}\frac{k}{2}\:\pm\:
                  i\,2\sqrt{1-\sin^{4}\frac{k}{2}}\;,
\end{equation}

\noindent
and components of the eigenvectors $E$ are given by

\begin{eqnarray}
\label{p7}
E_{2n-1}  &= &v(k)\,e^{ikn}
\nonumber\\
E_{2n}&= &u(k)\,e^{ikn}
\nonumber\\
E_{2n+1}&= &v(k)\,e^{ik(n+1)}
\nonumber\\
E_{2n+2}&= &u(k)\,e^{ik(n+1)}\,.
\end{eqnarray}

\noindent
Here $n = 1,3,\ldots,N-1$;  $k=\frac{2\pi}{N}m\ ,m=-(\frac{N}{2}-1),
\ldots,-1,0,1,\ldots,\frac{N}{2}$,
and amplitudes $u(k)$ and $v(k)$ are related by

\begin{equation}
\label{p8}
u_{\pm}= i\,e^{i\frac{k}{2}}(-\sin\frac{k}{2}\:\pm\:
            \sqrt{1+\sin^{2}\frac{k}{2}}\;\;)\,v_{\pm}(k)\,\,.
\end{equation}

\noindent
The normalization condition gives:

\begin{equation}
\label{p9}
\mid u_{\pm}\mid= \frac{1}{\sqrt{2N}}(1+\sin^{2}\frac{k}{2}\:\pm\:

\sin\frac{k}{2}\sqrt{1+\sin^{2}\frac{k}{2}}\;\;)^{-\frac{1}{2}}
\end{equation}
and
\begin{equation}
\label{p10}
\mid v_{\pm}\mid= \frac{1}{\sqrt{2N}}(1+\sin^{2}\frac{k}{2}\:\mp\:

\sin\frac{k}{2}\sqrt{1+\sin^{2}\frac{k}{2}}\;\;)^{-\frac{1}{2}}\;.
\end{equation}

Using eqs. (\ref{p6}) -- (\ref{p10}) it is possible to find the wave function
for any $t$. After a lengthy but straightforward calculation, the
even $n$ components are found to be

\begin{eqnarray}
\label{e10}
\psi_{n}(t) &= &\frac{1}{2^{t}}\sum_{m=-(\frac{N}{2}-1)}^{\frac{N}{2}}
               [\lambda_{+}^{t}(k_{m})\mid u_{+}(k_{m})\mid^{2}\,+\,
                \lambda_{-}^{\ast\,t}(k_{m})\mid u_{-}(k_{m})\mid^{2}\,]
               e^{ik_{m}\frac{n-N}{2}}\\
            &= &\frac{1}{N}\sum_{m=-(\frac{N}{2}-1)}^{\frac{N}{2}}
              [\cos(\varphi t)\cos(k_{m}\frac{n-N}{2})\,+\,
              \sin(\varphi t)\frac{\sin
k_{m}}{\sqrt{1+\sin^2\frac{k_{m}}{2}}}\,
               \sin(k_{m}\frac{n-N}{2})]\,\,,
\nonumber
\end{eqnarray}

\noindent
while the  odd $n$  components are given by

\begin{eqnarray}
\label{e11}
\psi_{n}(t) &= &-\frac{i}{2^{t}}\sum_{m=-(\frac{N}{2}-1)}^{\frac{N}{2}}
                [\lambda_{+}^{t}(k_{m})(\sin\frac{k_{m}}{2}+
                \sqrt{1+\sin^{2}\frac{k_{m}}{2}})\mid u_{+}(k_{m})\mid^{2}\,
\nonumber\\
            &  &+\,\lambda_{-}^{\ast\,t}(k_{m})(\sin \frac{k_{m}}{2}-
                 \sqrt{1+\sin^{2}\frac{k_{m}}{2}})\mid u_{-}(k_{m})\mid^{2}\,]
               e^{ik_{m}\frac{n-N}{2}}
\nonumber\\
            &= &\frac{1}{N}\sum_{m=-(\frac{N}{2}-1)}^{\frac{N}{2}}
               \frac{\sin(\varphi t)}{\sqrt{1+\sin^2\frac{k_m}{2}}}
                \cos(k_{m}\frac{n-N}{2})\,\,.
\end{eqnarray}

\noindent
In the above expressions
$\varphi = {\rm arg} (\lambda_{+})$.

Knowing the closed form for $\Psi(t)$ we can now analytically
calculate moments $\langle x\rangle$ and $\langle x^{2}\rangle$
as function of $t$.
We have found that
for large $t$

\begin{equation}
\label{e12}
\langle x^{n} \rangle= (1 - \frac{1}{\sqrt{2}})\,t^{n}\;,\;\;\;\;\;n=1,2\,.
\end{equation}

\noindent
Therefore,  we have explicitly reproduced  the well known result
that in the absence of disorder the wave packet spreads in time in a ballistic
way (quantum diffusion), i.e. $\sqrt{\langle x^{2}}\rangle\, \sim t$.
In the next section we shall use the same method to study the
propagation in an external "magnetic field".
\section{Interference effects due to phases induced by a "magnetic field"}
\label{sec:magnetic}

Previous studies [11-15] have addressed the tunneling of
strongly localized electrons for
which the "directed paths" is justified as an approximation.
In these studies a very
intriguing behavior was found due to the combined effects of the lattice and
the magnetic field on the interfering paths. Therefore we were compelled to
investigate how these combined effects will change in a unitary model, although
no direct relation to a realistic system exists for this model. We believe that
this study will add to the the general understanding on the behavior of lattice
electrons in a magnetic field.

   The "magnetic field" adds a phase to the bonds.
The phase $\varphi$
is determined by the applied magnetic field and it is given by the discretized
curvilinear integral of the vector potential along the bond between the two
sites. There are many possible choices for this phase which yield the same
magnetic field and which differ by a gauge transformation. Any gauge such that
the sum of all phases around a plaquette gives the correct
flux through the plaquette
(in units of the flux quantum $\Phi_{0}\,=\,\frac{\hbar}{e}$) can be used. Here
we use so called diagonal staggered
gauge used in \cite{SW2,MKSW}, $\varphi (t) =
\pm\alpha \frac{t}{2}$ with $\alpha = 2\pi\Phi/\Phi_{0}$ and $\Phi$ is the flux
per elementary plaquette. The main characteristic of this gauge is that phase
depends linearly on $t$ coordinate but there is no dependence on transverse
coordinate.

    Now we shortly describe the transfer matrix calculations in the
presence of a "magnetic field".
Each step  of propagation, say from  a site at $t$ to
a site at $t+1$ in our notation includes
two bonds, and therefore  two phases.
Let us call them $\varphi_{1}(t)$ and
$\varphi_{2}(t)$, and  denote their combinations by
$\varphi_{\pm}(t) = \varphi_{1}(t)\,\pm\, \varphi_{2}(t)$.
We  follow the procedure described in the previous section.
In the presence of  "magnetic field" the basic motif of
transfer matrix (\ref{p3})
depends on $t$ and it is given by:

\begin{equation}
\label{m1}
\begin{array}{rrrr}
-e^{-i\varphi_{-}}&-e^{-i\varphi_{-}} &-e^{ i\varphi_{+}} & e^{ i\varphi_{+}}
\\
 e^{-i\varphi_{+}}& e^{-i\varphi_{+}} &-e^{ i\varphi_{-}} & e^{ i\varphi_{-}}
\end{array}
\end{equation}

\noindent
In order to diagonalize $T(t)$, we again impose periodic boundary
conditions and use  the same eigenvectors as before.
Simple calculation gives corresponding eigenvalues:

\begin{equation}
\label{m2}
\lambda_{\pm}(k,t) = -2\sin(\frac{k}{2}-\varphi_{1}(t))\,\sin(\frac{k}{2}
                     -\varphi_{2}(t))\;
                     \pm\: i\,2\sqrt{1-\sin^{2}(\frac{k}{2}-\varphi_{1}(t))\,
                     \sin^{2}(\frac{k}{2}-\varphi_{2}(t))}\;.
\end{equation}

The presence of a  "magnetic field"  clearly breaks the standard procedure for
transfer matrix calculation.
Namely, transfer matrices at different $t$ do not commute
and therefore cannot be diagonalized simultaneously.
However, the product of $T$ matrices which has to be diagonalized can be
written
in a block-diagonal form with $2\times2$ matrices on diagonal.
Each $2\times2$ matrix depends on  a specific value of transverse
momentum vector
$k_{m}=\frac{2\pi}{N}\,m$. It is defined by a product of
$2\times2$ matrices at different $t$
but with the same value of $k$. Of course, in
the absence of magnetic
field, the off-diagonal elements vanish and diagonal elements are
simply $\lambda_{+} ^{t}(k)$ and $\lambda_{-}^{t}(k)$.
This suggests the possibility to use the straightforward transfer
matrix procedure together with  a perturbative treatment for small
magnetic field.
Perturbative results obtained in this way
are too long to be presented here.
In the next subsection we will approach the same problem
in a  different way.
\subsection{Mapping to the Ising model}
\label{sec:ising}

       We now discuss a different approach to the system
in presence of a "magnetic field".
The basic idea is to map our model to
an one-dimensional Ising model and then
to use the  transfer matrix techniques to find the partition function
which is related to the components of the wave function.

For future use it is now more
convenient to consider transfer over
each column of scatterers separately.
Thus, we assign a single unitary matrix modified
by the effect of magnetic field
to all sites with the
same $t$ (the sites along the same column):

\begin{equation}
\label{m3}
S = \left( \begin{array}{cc}
           \alpha_{n} e^{i\varphi_{n+1}} & \beta_{n} e^{i\varphi_{n+1}} \\
       -\beta_{n}^{*} e^{-i\varphi_{n+1}}& \alpha_{n}^{*} e^{-i\varphi_{n+1}}
    \end{array} \right)\,\,.
\end{equation}

\noindent
$\alpha$ and $\beta$ are any complex numbers and index $n=1,2,\ldots\;,t$
denotes position of
the scatterer along $t$-axis.
$\varphi_{n+1}$ is the change of phase of the wave function
during the time interval from $n$ to $n+1$.
It is positive (negative) for  the wave
propagating in  up (down) direction, and
for our gauge choice
it is a linear function of index $n$.

Again, the probability to reach point $(x,t)$ is obtained by
summing the individual amplitudes of all directed paths starting at
the origin and ending at  $(x,t)$.
To count all different paths  we
introduce a set of numbers $\sigma_{n}$
such that $\sigma_{n+1} =+1$
if the wave scattered at site $n$ goes through the upper bond
and $\sigma_{n+1} =-1$
if it goes through the lower bond, \cite{propagator}.
The sequence $\sigma_{1},\sigma_{2},\ldots\,,
\sigma_{t}$ uniquely specifies a path from the origin to
some site at distance $t$.
The final transversal position of a path is given by

\begin{equation}
\label{m4}
x = \sum_{n=1}^{t}\sigma_{n}\,.
\end{equation}

\noindent
The value of $\sigma_{0}$ is fixed by the initial condition:
$\sigma_{0} = +1 (-1)$  for the
incident wave coming through lower (upper) bond.
Physically,  $\sigma_{n}$ and $x$
can be identified with spins and
the magnetization of the system, respectively. The size of the system is
determined by $t$.
Keeping
this analogy in mind we will call components of the wave function
$\Psi$ simply the partition function $Z$.

Next, to each step between
the two sites we also
have to assign a factor depending on the scattering and
the "magnetic field". This can be
easily done for a path with a fixed spin configuration. For example, if
$\sigma_{n} =+1$ and $\sigma_{n+1} =+1$  we have
scattering from upper to
upper channel and corresponding factor is $\alpha\,e^{i\varphi_{n+1}}$.
Combination $\sigma_{n} =+1,\;\sigma_{n+1} =-1$
corresponds to scattering from upper to
lower channel which is described by $-\beta^{*}\,e^{-i\varphi_{n+1}}$.
Similarly, for
$\sigma_{n} =-1,\;\sigma_{n+1} =+1$ we have a factor
$\beta\,e^{i\varphi_{n+1}}$ and for
$\sigma_{n} =-1,\;\sigma_{n+1} =-1$ a factor $\alpha^{*}\,e^{-i\varphi_{n+1}}$.

The wave amplitude (partition function) to reach the site $(x,t)$ with initial
condition given by the value of $\sigma_{0}$ is

\begin{eqnarray}
\label{m5}
Z(x,t) &= &\sum_{\sigma_{1} \ldots \sigma_{t}}^{}\!'\prod_{n=0}^{t-1}
    [\alpha_{n}e^{i\varphi_{n+1}}]^{\frac{1}{4}(1+\sigma_{n})(1+\sigma_{n+1})}
            [\beta_{n} e^{i\varphi_{n+1}}]^{\frac{1}{4}(1-\sigma_{n})
            (1+\sigma_{n+1})}
\nonumber \\
       &  & [-\beta_{n}^{*} e^{-i\varphi_{n+1}}]^{\frac{1}{4}(1+\sigma_{n})
            (1-\sigma_{n+1})}[\alpha_{n}^{*}
            e^{-i\varphi_{n+1}}]^{\frac{1}{4}(1-\sigma_{n})(1-\sigma_{n+1})}
\nonumber \\
       &= &\sum_{\sigma_{1} \ldots \sigma_{t}}^{}\!'
           e^{i\,\sum_{n=1}^{t}\,\varphi_{n}\sigma_{n}}\prod_{n=0}^{t-1}
    exp[\,\sigma_{n}\sigma_{n+1}\,(\frac{1}{2}\ln\frac{|\alpha|}{|\beta|}
        - i\,\frac{\pi}{4})
        +  \frac{1}{2}\ln|\alpha \beta| + i\,\frac{\pi}{4}
\nonumber \\
       &  &+  i\,\sigma_{n}(\frac{arg(\alpha) - arg(\beta)}{2} + \frac{\pi}{4})
    +\;\; i\,\sigma_{n+1}(\frac{arg(\alpha)+arg(\beta)}{2}
-\frac{\pi}{4})\,]\;,
\end{eqnarray}

\noindent
where the prime on
the summation sign indicates that the partition function $Z$
has to be evaluated  for constant magnetization, i.e.
under the constraint $\sum_{n=1}^{t}\sigma_{n} = x$.

   In terms of $\alpha$ and $\beta$ expression (\ref{m5}) is still very general
and  can be used to study layered disorder,
i.e. the problem when all scatterers along the same
$t$ are equivalent to each other but they might be
totally independent from scatterers
at different $t$.
For dissipationless wave propagation the normalization condition requires
$|\alpha|^{2} + |\beta|^{2} = 1$. If magnitudes of the $\alpha$ and $\beta$
are equal, then a wave is scattered into upper or lower channel with equal
probability. Randomness in the ratio $|\alpha|/|\beta|$ will lead to a
"random bond" in the one dimensional Ising model
(this will be clarified below). On the
other side, randomness in arguments of $\alpha$ and $\beta$ could be
interpreated as a random field Ising model.
Our goal here is to study new features which come from the "magnetic field"
alone.
Thus, we can choose a
simple form for matrix  elements $\alpha$ and $\beta$, for example

\begin{equation}
\label{m6}
 \alpha = \frac{1}{\sqrt{2}}\:\:\:\:\:\:\:\:
 \beta  = \frac{i}{\sqrt{2}}\:.
\end{equation}

\noindent
The choice (\ref{m6})  yields

\begin{equation}
\label{m8}
Z(x,t) = (\frac{1}{\sqrt{2}})^{t}\sum_{\sigma_{1} \ldots \sigma_{t}}{}'
  e^{[\;J\sum_{n=0}^{t-1}(\sigma_{n}\sigma_{n+1}\,-\,1)\,+\,i\,\sum_{n=1}^{t}\,
     \varphi_{n}\sigma_{n}\;]}
\end{equation}

\noindent
where $J$ is defined by relation $e^{-2J}=i$ or $J=-i\frac{\pi}{4}$.

The constrained sums make further progress very difficult.
To get rid of this
constraint it is convenient to introduce the momentum representation by

\begin{eqnarray}
\label{m9}
Z(k,t) &= &\frac{1}{N}\sum_{x}Z(x,t)\,e^{-ikx}
\nonumber \\
       &= &\frac{1}{N}(\frac{1}{\sqrt{2}})^{t}\sum_{\sigma_{1} \ldots
           \sigma_{t}}^{}
  e^{[\;J\sum_{n=0}^{t-1}(\sigma_{n}\sigma_{n+1}\,-\,1)\,+\,i\,\sum_{n=1}^{t}\,
     h_{n}\sigma_{n}\;]}
\end{eqnarray}

\noindent
where $k$ is one of the  $k_{m}=\frac{2\pi}{N}m$ and $h_{n} = \varphi_{n}- k$.
It is clear from eq. (\ref{m9})
that a  Fourier transform of the amplitude to reach
point $(x,t)$ is precisely the partition function for
a generalized one-dimensional
Ising model.
Compared to the ordinary Ising model
we have now a complex coupling constant $J$ and
the local magnetic field $h_{n}$ has an imaginary prefactor.
The standard Ising model  in inhomogeneous
magnetic field was
studied in ref. \cite{DFP}.
In our case the complex exponent will lead to
some divergences.  We shall  identify
these divergences and give a prescription how to handle them.

    The partition function in $k$-space can be written in the form

\begin{equation}
\label{m10}
Z(k,t) = Z_{+}(k,t) + Z_{-}(k,t)
\end{equation}
\noindent
where

\begin{equation}
\label{m11}
Z_{\pm}(k,t) = (\frac{1}{\sqrt{2}})^{t}\sum_{\sigma_{1} \ldots
\sigma_{t-1}}^{}
     e^{[\;J\sum_{n=0}^{t-2}(\sigma_{n}\sigma_{n+1}\,-\,1)\,+\,
     i\,\sum_{n=1}^{t-1}\,h_{n}\sigma_{n}\;+ J(\pm \sigma_{t-1}-1)\;
     \pm\;i h_{t}\;]}\,\,.
\end{equation}

\noindent
$Z_{+}$ and $Z_{-}$ are the $k$-space amplitudes to reach point
$(x,t)$ from the
upper or lower  direction, respectively.
It is easy to see that the following recursive
relations hold:

\begin{equation}
\label{m13}
Z_{\pm}(n) = \frac{1}{\sqrt{2}} e^{\pm ih_{n}}\,
             [\;Z_{\pm}(n-1) + e^{-2J}\,Z_{\mp}(n-1)\;]\,\,.
\end{equation}

\noindent
Defining the two ratios $r_{+}(n)$ and $r_{-}(n)$  by

\begin{equation}
\label{m15}
r_{\pm}(n) = \frac{Z_{\mp}(n)}{Z_{\pm}(n)}
           = e^{\mp 2ih_{n}}\,
                  \frac{r_{\pm}(n-1)+e^{-2J}}{e^{-2J}r_{\pm}(n-1)+1}
\end{equation}

\noindent
after some lengthy but straightforward algebra we find

\begin{equation}
\label{m16}
Z_{\pm}(k,t) = \frac{1}{N}(\frac{1}{\sqrt{2}})^{t}\,e^{J(\pm \sigma_{0}-1)}
                \,e^{\mp ikt}\, e^{\pm i\sum_{n=1}^{t}\varphi_{n}}
                \; \prod_{n=1}^{t-1}\!'\,[\,1 + i\,r_{\pm}(n)\;]\;.
\end{equation}

\noindent
The prime on the product sign means that care must be taken
when the ratios $r_{\pm}$ diverge.
In what follows we describe how to treat this problem.
We start with

\begin{equation}
\label{m17}
 r_{\pm}(n=1) = e^{\mp2J\sigma_{0}}
                e^{\mp 2i(\varphi_{1}-k)}
\end{equation}

\noindent
and use eq. (\ref{m15})
to evaluate ratios for larger values of $n$.
If for some $n$, $r_{\pm}(n) = e^{-2J} = i$
 (yielding a divergent term) then $(n+1)^{\rm th}$ term in the product
is set to one and  the
next two terms in the product are given by

\begin{equation}
\label{m18}
 r_{\pm}(n+2) = e^{2J}\,[\,(e^{-4J}-1)e^{\mp 2i(\varphi_{n+2}-k)}\,-\,1\;]
\end{equation}
and
\begin{equation}
\label{m19}
 r_{\pm}(n+3) = e^{2J}\,e^{\mp 2i(\varphi_{n+3}-k)}\;.
\end{equation}

\noindent
The contributions of terms given by eqs. (\ref{m18}) and (\ref{m19})
to the product in eq.  (\ref{m16}) are
\begin{equation}
\label{m20}
 1 + i\;r_{\pm}(n+2) = -2\;e^{\mp 2i(\varphi_{n+2}-k)}
\end{equation}
and
\begin{equation}
\label{m21}
 1 + i\;r_{\pm}(n+3) = 2\cos(\varphi_{n+3}-k) e^{\mp i(\varphi_{n+3}-k)}\;.
\end{equation}

The iterative procedure described above gives Fourier transform of $Z$.
Returning to the real space by an inverse transform

\begin{equation}
\label{m22}
Z_{\pm}(x,t) = \sum_{k}Z_{\pm}(k,t)\,e^{ikx}\;,
\end{equation}

\noindent
and computing the probability for a wave to end at  point $(x,t)$
we find

\begin{equation}
\label{m23}
\mid Z(x,t)\mid^{2} = \mid Z_{+}(x,t)\mid ^{2} +  \mid Z_{-}(x,t)\mid ^{2}\;.
\end{equation}
\subsection{Numerical results and discussion}
\label{sec:numeric}

     The expression (\ref{m16}) together with eqs. (\ref{m17})-(\ref{m19})
uniquely determine  the probability distribution after $t$ steps
for any magnetic
flux. As is pointed out in ref. \cite{SW2}, its behavior strongly depends
on the commensurability of $\alpha = 2\pi\Phi/\Phi_{0}$.
If $\alpha = p/q$, ($p,q$ are integers), the calculation of the
partition
function (\ref{m9}) can be related to the case when magnetic flux is not
present. To show  it, we  write eq. (\ref{m9}) in the form

\begin{equation}
\label{n1}
Z(k,t) = \frac{1}{N}(\frac{1}{\sqrt{2}})^{t}\sum_{\sigma_{1} \ldots
\sigma_{t}}{}
T(\sigma_{0},\sigma_{1}) T(\sigma_{1},\sigma_{2})\cdots\;
T(\sigma_{t-1},\sigma_{t})\:e^{\frac{i}{2}(\sigma_{t}h_{t} - \sigma_{0}h_{0})}
\end{equation}

\noindent
where $T(\sigma_{n},\sigma_{n+1})$ is $2\times2$ transfer matrix

\begin{equation}
\label{n2}
T(\sigma_{n},\sigma_{n+1}) =
e^{J(\sigma_{n}\sigma_{n+1}-1) + \frac{i}{2}(\sigma_{n}h_{n}
+\sigma_{n+1}h_{n+1})}.
\end{equation}

\noindent
The extra field $h_{0}$ is defined because of symmetry and it is set to be
zero.
The transfer matrices in eq. (\ref{n1}) do not commute.
This is a  consequence of their dependence on the
magnetic flux (index $n$ is related to our gauge choice). However, for rational
$\alpha$ the $n$-dependence can be eliminated.
Grouping transfer matrices in eq. (\ref{n1}) into groups with $q$ factors we
will get a new set of $n/q$ commuting matrices
$T' = T(\sigma_{n},\sigma_{n+1})\cdots\;T(\sigma_{n+q},\sigma_{n+q+1})$,
which is $n$-independent. The grouping procedure
corresponds to rescaling our initial lattice with coordinates $(x,t)$ to
the new lattice $(x',t')$ so that $x=qx'$ and $t=qt'$. On rescaled lattice
effect of an external field is apparently eliminated and problem can be
solved as in Sec. 3.

   Diagonalizing matrix $T'$ yields two eigenvalues $\lambda_{1,2}$ and
corresponding eigenvectors $E_{1,2}$ which are orthogonal and normalized:

\begin{equation}
\label{n3}
  E^{*}_{i}(+)E_{j}(+) + E^{*}_{i}(-)E_{j}(-) = \delta_{i,j}
\end{equation}

\noindent
where +/- denotes upper/lower component of the eigenvector $E_{i}$. In terms
of the eigenvalues and the eigenvectors, the partition function is

\begin{equation}
\label{n4}
 Z(k,t) = \frac{1}{N}(\frac{1}{2})^{\frac{t}{2}}\;\sum_{\sigma_{t}}{}
      \{\;\lambda^{\frac{t}{q}}_{1} E^{*}_{1}(\sigma_{0})E_{1}(\sigma_{t})
       +  \lambda^{\frac{t}{q}}_{2} E^{*}_{2}(\sigma_{0})E_{2}(\sigma_{t})\;\}
          \; e^{\frac{i}{2}(\sigma_{t}h_{t} -\sigma_{0}h_{0})}.
\end{equation}

\noindent
Eq.(\ref{n4}) has the same structure as eqs. (\ref{e10}) and (\ref{e11})
but, as we have argued, it also describes the effect of rational magnetic flux.
 In Fig. 2a the full line  shows probability distribution for the case
when the applied magnetic flux is $\alpha=1/3$.
For this case the matrix $T'$ is built from three $T$ matrices in sequence.
The eigenvalues in eq. (\ref{n3}) are:

\begin{equation}
\label{n7}
 \lambda_{1,2} = -2 \sin(k)\;(\cos(\frac{\pi}{3}) + 2 \cos^{2}(k)\;)
   \pm 2i \sqrt{2 - \sin^{2}(k)[ \cos(\frac{\pi}{3}) + 2 \cos^{2}(k)\;]^{2} }
\end{equation}

\noindent
and corresponding eigenvectors

\begin{equation}
E_{1,2} \sim \left( \begin{array}{c}
-e^{-ik}(\sin(2k) - i\;\cos(\frac{\pi}{3}) ) - \sqrt{2}e^{\pm i\varphi}  \\
    -\cos(2k) + i\;\sin(\frac{\pi}{3}) \end{array} \right)
\end{equation}

\noindent
where $\varphi = arg(\lambda_{+})$.

    From Figs. 2a and 2b we see that increasing "magnetic flux"
causes the beam to
become more collimated. The general form of the probability
distribution remains unchanged.
This is precisely  what one would expect following the above
arguments. Namely, introducing magnetic flux we effectively rescale our
lattice  to some smaller lattice $(x',t')$. Smaller $t$ leads to smaller
diffusion. Diffusion of the propagating wave as function of $t$ is
discussed in Sec. 3.

   The behavior of the probability distribution is very sensitive on value of
$\alpha$. Even small changes from rational $\alpha$ to some close irrational
number produces dramatical effects. In Fig. 2a and 2b probability
distributions after $t=100$ steps are compared for
$\alpha = 1/3 =0.333..., \frac{1}{\sqrt{8}}=0.353...$
and $\alpha = 0.6,\frac{\sqrt{5}-1}{2}=0.618...$. For irrational $\alpha$, the
distribution is concentrated to the small region around $x=0$.
The  probability to
reach point $x=0$ as function of $t$ is shown in Fig. 3 for rational and
irrational $\alpha$. While for rational $\alpha$ $P(x=0,t)$ decays with
regular oscillations, irrational $\alpha$ produces aperiodic structure
which does not decay with $t$.
The high sensitivity on irrational $\alpha$ could be
traced back to eq. (\ref{m21}) giving the product of
$\cos(\pi\alpha n)$ factors.
Recently, in ref. \cite{FSW}, the behavior of the localized
electrons on a lattice with incommensurate magnetic flux was investigated.
The general features we find here are similar to those derived in \cite{FSW}.

\section{Directed waves in strongly disordered media}
\label{sec:disorder}

In this section we discuss the wave propagation in a random medium. We describe
the randomness by taking the scattering matrix at each lattice site to be an
independent random element of the $U(2)$ group. We are interested in the
behavior of the  probability distribution and its higher moments as well as the
transverse moments of $x$, i.e., $[\langle x^{n} \rangle ]$ and
$[\langle x \rangle^{n}]$ as functions of the longitudinal distance $t$. To
calculate
$[\langle x \rangle ^{n}]$ one needs to know the  correlations among $n$ paths
propagating on the lattice with a given realization of randomness. These
correlations are the  primary subject of our interest.
Using some simple formulae
for group integration we will develop a systematic and exact procedure for
averaging various quantities of physical interest over the disorder.

   The probability $P(x,t)$ to reach a given point $(x,t)$ can be
calculated by summing over all possible paths which begin at the origin and
end at $(x,t)$.  Denoting  by $A_{i}(x,t)$ the contribution of the $i^{\rm th}$
path, the probability  $P(x,t)$ is given by

\begin{equation}
\label{d1}
P(x,t) = \mid \sum_{i}^{}A_{i}(x,t) \mid ^{2}
       = \sum_{i_{1},i_{2}}^{} A_{i_{1}}^{*}(x,t)\,A_{i_{2}}(x,t)\;.
\end{equation}

\noindent
The amplitude $A_{i}$ is simply the product of $t$ random, mutually independent
elements of the $S$-matrices along the $i^{\rm th}$ path.  All amplitudes are
defined for the same realization of the disorder. The contributions of two
paths
$i_1$ and $i_2$ to the sum in (\ref{d1}) depend on their mutual relation.
The two paths can be totally disconnected (meeting only at the end points),
they can overlap along some links or intersect at some lattice sites.
On their common
parts they share the same factors.  From a statistical point of view, the most
interesting phenomena are related to intersections and merging or branching
points.  At these points paths do not have assigned the same factors but
factors
from the same scattering matrices.

   We now turn to the question of the statistical averaging over the quenched
disorder. We assume  a uniform distribution of $S$ matrices over all unitary
matrices. Hence,  with an equal probability the scattering matrix may be
any $U(2)$ matrix. In this case there is no difference between averaging over
either the $U(2)$ or the   $SU(2)$ group.  Other choices, such as distributions
preferring some
propagating  direction are also possible, but will not be  discussed here.

   The quenched average of the probability distribution $P$ over all
realization
of the disorder is straightforward to compute.  After averaging, only the
paired or
neutral paths (such that one  member is selected from $A_{i}^{*}$ and the other
from $A_{i}$) give nonzero contribution to the sum in  eq. (\ref{d1}). Using
the
following integral over the $SU(N)$ group \cite{Cm}

\begin{equation}
\label{d2}
\int dS\,S_{ij}S_{lk}^{*} = \frac{1}{N} \delta_{jk}\delta_{il}
\end{equation}

\noindent
with $N=2$ and
counting  the number
of different paths from the origin to the point $(x,t)$ we obtain

\begin{equation}
\label{d3}
[P(x,t)] =  \frac{1}{2^{t}}\frac{t!}{(\frac{t+x}{2})!(\frac{t-x}{2})!}\:.
\end{equation}

\noindent
This formula is all what is needed to calculate all transverse moments
$[\langle x^{n}\rangle]$.  We are particularly interested in the second moment
characterizing the beam position. In this case, eq. (\ref{d3}) gives

\begin{equation}
\label{d4}
[\langle x^{2}(t) \rangle] = \sum_{x} x^{2} [P(x,t)] = t\;.
\end{equation}

\noindent
The result (\ref{d4}) was already derived in \cite{SKR,OE} and is also in
agreement with our numerical simulations shown in Fig. 5. It describes the
ordinary diffusion of a classical particle in a stochastic medium. Classical
diffusion is naturally expected since in eq. (\ref{d3}) there is no
contribution
from intersecting paths.  Comparing with the pure case (\ref{e12}), we conclude
that averaging over randomness removes the interference effects in $P$. The
probability distribution $P$ for propagation over nonrandom and random
scatterers is shown in Fig. 4. For the random case averaging was performed on
500 samples. By increasing the number of samples curve becomes smoother showing
no effect of interference.

   Nontrivial problems arise if one tries to find the scaling properties of the
beam center $[\langle x \rangle ^{2}]$. Results of long time Monte Carlo
simulations are shown in  Fig. 5.  Our numerical results are consistent with
those given in \cite{MKS} but the value of the scaling exponent is still
unclear.  Computer simulations with larger $t$ and more samples could help to
resolve this uncertainty.  However, Saul, Kardar and Read \cite{SKR} suggested
a way to reduce numerical difficulties. They proposed  to construct recursion
relations  among exactly averaged quantities at different $t$.  We will follow
their approach.

   The basic idea is to calculate the correlation functions averaged over the
disorder. They are  defined by

\begin{equation}
\label{d5}
W_{n}(x_{1},\ldots\, ,x_{n},t) = [P(x_{1},t)\ldots\,P(x_{n},t)]\;.
\end{equation}

\noindent
For $n=1$, $W_{1}(x,t)$ is given by (\ref{d1}). It describes  the time
evolution of one neutral path made up of a segment going from origin to $(x,t)$
and its complex conjugate which can be viewed as the time evolution in the
opposite direction. For $n\geq 2$ eq. (\ref{d5}) gives the probability that
$n$ neutral paths starting at the origin will end at position
$(x_{1},x_{2},\ldots\, ,x_{n})$ at distance $t$.
Some characteristic configurations
for time evolution of $W_{2}, W_{3}$ and $W_{4}$ are shown in Fig. 6. $W_{n}$
must satisfy  the initial conditions

\begin{equation}
\label{d6a}
           W_{n}(x_{1},x_{2},\ldots\, ,x_{n},t=0) =
           \delta_{x_{1},0}\delta_{x_{2},0}\ldots \delta_{x_{n},0}\;,
\end{equation}

\noindent
as well as a  sum rule which is a consequence of the probability conservation:

\begin{equation}
\label{d6}
\sum_{x_{1}}{} \sum_{x_{2}}{}\cdots \sum_{x_{n}}
           W_{n}(x_{1},x_{2},\ldots\, ,x_{n},t) = 1\;.
\end{equation}

\noindent
The role of $W_{n}$ in calculation of $[\langle x\rangle^{n}]$ is seen from the
following relation

\begin{equation}
\label{d7}
[\langle x\rangle^{n}] = \sum_{x_{1}}{} \sum_{x_{2}}{}\cdots \sum_{x_{n}}
      x_{1}x_{2}\cdots\, x_{n}\,[P(x_{1},t)P(x_{2},t)\cdots\, P(x_{n},t)] \;.
\end{equation}

    Evaluation of $W_{n}$ is based on the recursion relations connecting
$W_{n}$
at distance  $t$ and $t+1$. The recursion relations can be
derived by using symmetry
arguments and recalling that only neutral paths can survive disorder averaging.
The neutral paths, however, may cross each other and exchange partners as in
Fig. 6. The exchange effect means that we are dealing with interacting paths.
It is convenient  at this point to introduce relative coordinates
$(x,r_{1},r_{2},\ldots\,r_{n-1})$ instead of the previous set
$(x_{1},x_{2},\ldots\,x_{n})$. The coordinate $r$ is distance between paths at
some specified $t$ (see Fig. 6). The calculation of $W_{2}(x,r)$ is relatively
simple. First we look for all different ways to construct $W_{2}(x,r,t+1)$ from
$W_{2}(x,r,t)$.  Neglecting the crossings at $t$ there are four
possibilities as is shown in Fig. 7.  $W_{2}(x,r,t+1)$ is therefore  a
combination of $W_{2}(x\pm1,r,t)$ and $W_{2}(x\pm1,r\pm1,t)$. Next step is to
include the effects of paths exchange. They are present only if $r=0,\pm1$ at
$t$. This leads to the following terms in the recursion relation:

\begin{equation}
\label{d8}
\frac{1}{4}(1+\epsilon_{0}\delta_{r,0})W_{2}(x\pm1,r,t)
\end{equation}

\begin{equation}
\label{d9}
\frac{1}{4}(1+\epsilon_{1}\delta_{r,\mp1})W_{2}(x\pm1,r\pm1,t)
\end{equation}

\noindent
where we have parametrized the effect of disorder by $\epsilon_{0}$ and
$\epsilon_{1}$. Due to the symmetry $r \rightarrow -r$, the coefficients of
$W_{2}(x+1,r+1,t)$ and $W_{2}(x-1,r-1,t)$ must be equal.  This observation,
although trivial for $W_{2}$, is very useful for calculation of $W_{n}$ for
higher $n$.The sum rule (\ref{d6}) implies that the sum of all
the coefficients must be equal to one yielding $\epsilon_{0} = -\epsilon_{1}$.
Thus, the  recursion relation for $W_2$ can be written in  the form:

\begin{eqnarray}
\label{d10}
W_{2}(x,r,t+1) &= &
\frac{1}{4}(1+\epsilon\delta_{r,0})\{\;W_{2}(x+1,r,t) + W_{2}(x-1,r,t) \;\}+
\\
               &  &\frac{1}{4}(1-\epsilon\delta_{r,1})W_{2}(x-1,r-1,t) +
                   \frac{1}{4}(1-\epsilon\delta_{r,-1})W_{2}(x+1,r+1,t)\:.
\nonumber
\end{eqnarray}

The left-hand side of eq. (\ref{d10}) may be directly calculated by averaging
the quantity \mbox{$[P(x_{1},t+1)P(x_{2},t+1)]$} over the $SU(2)$ group.
This is easily done for small
$t$ since the number of possible configurations which must be taken into
account
is relatively small. In addition to the result (\ref{d2}), we also need a
formula for an integral over four $SU(2)$ elements \cite{Cm}

\begin{eqnarray}
\label{d11}
\int dS\,S_{ij}S_{lk}^{\ast}S_{mn}S_{qp}^{\ast} &= &
     \frac{1}{N^{2}-1}(\delta_{il}\delta_{mq}\delta_{jk}\delta_{np} +
                       \delta_{iq}\delta_{ml}\delta_{jp}\delta_{nk}) +
\\
\nonumber
&  &         \frac{1}{N(N^{2}-1)}(\delta_{il}\delta_{mq}\delta_{jp}\delta_{nk}
+
                       \delta_{iq}\delta_{ml}\delta_{jk}\delta_{np})\:.
\end{eqnarray}

\noindent
It turns out that $\epsilon = \frac{1}{3}$. The same procedure can be used for
a systematic evaluation  of other correlation functions. In the Appendix
we give the results for $W_{3}$ and $W_{4}$.

   Now we investigate the  $t$-scaling of $W_{n}(x=r_{1}=\cdots\;=x_{n-1}=0,t)$
and $\sum_{x}^{}W_{n}(x,r_{1}=r_{2}=\cdots\;=r_{n-1}=0,t)$. In the language of
random walk the former quantity is the probability that $n$ paired paths
reach the same point $x=0$ after time $t$ and  the latter is the probability of
their
reunion somewhere (see eq. (\ref{d5})). The defining equation of $W_{n}$,
(\ref{d5}), gives

\begin{equation}
\label{d12}
\sum_{x}^{}W(x,r_{1}=r_{2}=\cdots\;=r_{n-1}=0,t)
=\sum_{x}{}[P(x,t)^{n-1}P(x,t)]
\end{equation}

\noindent
which are exactly the moments of the probability
distribution itself. Recently it
was argued by Bouchaud et al. \cite{BTS} that the evolution of the wave must be
described by an infinite set of exponents defined by

\begin{equation}
\label{d13}
\sum_{x}^{}[P(x,t)] \approx t^{\mu(n)}\:.
\end{equation}

\noindent
The nontrivial form of the function $\mu(n)$ is a sign of ``multifractal"
structure of the propagating wave. For $n = 2,3,4$ the behavior of the $\mu$
can be found simply by iterating the recursion relations (\ref{d5}), (\ref{a1})
and (\ref{a2}) numerically. The numerical results shown in Fig. 8 and Fig. 9
suggest that for large $t$

\begin{eqnarray}
\label{d14}
W_{n}(x=r_{1}=\cdots\;=r_{n-1}=0,t))            &\sim & t^{-\frac{n}{2}}
\\
\label{d15}
\sum_{x}^{}W_{n}(x,r_{1}=\cdots\;=r_{n-1}=0,t)) &\sim &
                                            t^{-\frac{n}{2} + \frac{1}{2}}\:.
\end{eqnarray}

\noindent
These relations indicate that for large $t$ the interaction among paths is
irrelevant \cite{Fm}.

    The recursion relation (\ref{d10}) for $W_{2}(r=0,t)$ can be
diagonalized for small $\epsilon$. The nonperturbed eigenvectors are $\sin$
or $\cos$ functions and matrix elements of $\epsilon$-dependent terms in
eq. (\ref{d10}) are

\begin{equation}
\label{d16}
V_{n.m} = \delta_{n,0}\{\; \frac{1}{2} \epsilon\delta_{m,o} -
          \frac{1}{4}\epsilon\delta_{|m|,1}\;\}.
\end{equation}

\noindent
The eigenvalues of the transfer matrix relating $W_{2}(r=0,t+1)$ and
$W_{2}(r=0,t)$ are

\begin{equation}
\label{d17}
\lambda_{\pm}(k_{m}) = \cos^{2}(\frac{k_{m}}{2})\{ 1 \pm
        \epsilon\;\frac{4}{2N+1}\sin^{2}(\frac{k_{m}}{2})\;\} + O(\epsilon^{2})
\end{equation}

\noindent
where $k_{m}= \frac{2\pi}{2N+1}m,\; m=0,\pm 1,\ldots\;\pm N$. For large $t$,
the eigenvalues
(\ref{d17}) and corresponding eigenvectors yield

\begin{equation}
\label{d18}
W_{2}(r=0,t) = \frac{1}{2^{2t}} \left( \begin{array}{c} 2t\\t
\end{array}\right)
             + O(\frac{\epsilon}{\sqrt{t}}).
\end{equation}
\noindent
Equation (\ref{d18}) shows that the effect of $\epsilon$ is too local to
produce
any relevant effect for large $N$. This supports the predicted scaling behavior
in eqs (\ref{d14}) and (\ref{d15}).

\section{Conclusions}
\label{sec:conclusion}

   In this work we studied different aspects of a lattice model for unitary
propagation. We first presented a closed-form solution for such a propagation
in a uniform medium. Then we took a theoretical "detour" to study the case when
the phases were modulated as if they were due to an external "magnetic field".
As emphasized before this situation does not describe lattice electrons with a
field in 1D or 2D. It is rather an interesting theoretical extension for
the "directed paths" model of localized lattice electrons in 2D, which can
widen our general understanding of this important class of systems.
For the non-unitary model, the complete
solution was found, \cite{SW2,MKSW}. Interesting inflation rules were found for
$\Phi/\Phi_{0} = p/q$. For incommensurate $\Phi/\Phi_{0}$ an aperiodic
behavior with striking scaling properties was revealed \cite{FSW}.
In the unitary case studied
here, we have shown that the model cannot be solved completely by
straightforward application of transfer matrix techniques but we managed to
find a closed form for the probability distribution of propagating waves by
mapping our problem to a generalized Ising model with imaginary coupling
constant $J$ and imaginary inhomogeneous external field. The
problem was also investigated numerically.
Inflations rules for rational
$p/q$ hold in this case too but they are certainly not as trivial. The wave
function amplitude becomes more and more "collimated" around $x=0$ as $q$ is
increased.

    We next studied the disordered case. Because of the "light cone" constraint
on the lattice we have found $\langle x^{2} \rangle \sim t$, while the
behavior off-lattice is known to be $\langle x^{2} \rangle \sim t^{3}$,
\cite{JK}.
So in presence of disorder the present lattice model does not recover the
continuum limit. The continuum analysis was performed on the Schr\"{o}dinger
equation for a particle moving in a random potential which changes rapidly
both in space and in time.
Our results for $[\langle x\rangle^{2}]$ agree with that of Saul et al.
\cite{SKR}.

We have also investigated the higher moments of $P(x,t)$.
Our numerical investigations were based on the recursion relations between
various moments. It should be emphasized that the averaging over disorder
leads to exact (but cumbersome) relations that are then iterated numerically
(the iteration process does not involve randomness anymore). Our numerical
results suggest a simple "gap scaling" $[P^{n}(x,t)] \sim t^{-\frac{n}{2}}$.
A simple analytic argument explains this behavior for $n = 2$, (see also
\cite{SKR}). Although we cannot prove it rigorously, it looks very likely
that these analytic arguments hold for $n > 2 $ as well. How to reconcile
this "gap scaling" with the multifractal behavior found by Bouchaud et al?
In their model the wave function is defined on the sites and unitarity was
ensured by replacing the evolution operator
$e^{-\frac{i}{\hbar}\hat{{\cal H}}t}$ by the Cayley operator
\mbox{$(1 - \frac{i}{\hbar}\hat{{\cal H}}t)/
(1 + \frac{i}{\hbar}\hat{{\cal H}}t)$}.
So the most likely possibility that comes to mind is that the behavior is very
sensitive to the type of latticization used. The fact that neither lattice
formulation yields the correct continuum limit leaves open the unusual
situation in which different lattice models may belong to different
universality classes. It should be also pointed out that the deviations from
"gap scaling" found in \cite{BTS} are very small
for these positive moments but are substantial for negative moments which are
not studied here. More investigations of this and related issues will be
most worthy.

\acknowledgments

We are grateful to M. Kardar for useful conversations and critical reading of
the manuscript. We are also thankful to J.P. Bouchaud and E. Medina for
clarifying comments.  Acknowledgement is made to the donors of the Petroleum
Research Fund, administrated by the ACS, for support of this research.

\appendix
\section*{Recursion relations for $W_{3}$ and $W_{4}$}

Here we present the recursion relations
for the correlation functions
$W_{3}(x,r_{1},r_{2},t)$ and $W_{4}(x,r_{1},r_{2},r_{3},t)$.
The derivation is
completely analogous to the one described in Sec. 5 for $W_{2}(x,r,t)$.
However, for given coordinates of the ending point at distance $t$ there are
now
many more path  configurations contributing to $W$ which make the calculation
somewhat tedious.  For  $W_{3}$  one finds

\begin{eqnarray}
\label{a1}
W_{3}(x,r_{1},r_{2},t+1) &= &\frac{1}{8}\sum_{a}{}
                             W_{3}(x+a_{0},r_{1}+a_{1},r_{2}+a_{2},t)
\\
                         &  &
     \{ 1 + \frac{1}{3}[
    (-1)^{a_{1}}\delta_{r_{1}+a_{1},0} +
    (-1)^{a_{2}}\delta_{r_{2}+a_{2},0} +
    (-1)^{a_{1}+a_{2}}\delta_{r_{1}+a_{1},r_{2}+a_{2}}]\,\}
\nonumber
\end{eqnarray}
where the sum is over $a = (a_{0},a_{1},a_{2}) = (\pm1,0,0),(\pm1,\pm1,0),
(\pm1,0,\pm1),(\pm1,\pm1,\pm1)$.
Similarly the recursion relation for $W_{4}$ reads

\begin{eqnarray}
\label{a2}
W_{4}(x,r_{1},r_{2},r_{3},t+1) &=& \frac{1}{16}\sum_{a}{}
                  W_{4}(x+a_{0},r_{1}+a_{1},r_{2}+a_{2},r_{3}+a_{3},t)\;
\\
          &  &
    \{ 1 + \frac{1}{3}[
    (-1)^{a_{1}}\delta_{r_{1}+a_{1},0} +
    (-1)^{a_{2}}\delta_{r_{2}+a_{2},0} +
    (-1)^{a_{3}}\delta_{r_{3}+a_{3},0} +
\nonumber\\
          &  &
    (-1)^{a_{1}+a_{2}}\delta_{r_{1}+a_{1},r_{2}+a_{2}} +
    (-1)^{a_{1}+a_{3}}\delta_{r_{1}+a_{1},r_{3}+a_{3}} +
\nonumber\\
          &  &
    (-1)^{a_{2}+a_{3}}\delta_{r_{2}+a_{2},r_{3}+a_{3}} ] +
     \frac{1}{9}(-1)^{a_{1}+a_{2}+a_{3}}[
\nonumber\\
          &  &
     \delta_{r_{1}+a_{1},0}\delta_{r_{2}+a_{2},r_{3}+a_{3}} +
     \delta_{r_{2}+a_{2},0}\delta_{r_{1}+a_{1},r_{3}+a_{3}} +
     \delta_{r_{3}+a_{3},0}\delta_{r_{1}+a_{1},r_{2}+a_{2}} ]
\nonumber\\
          &  &
     -\frac{2}{15}(-1)^{a_{1}+a_{2}+a_{3}}
     \delta_{r_{1}+a_{1},0}\delta_{r_{2}+a_{2},0}\delta_{r_{3}+a_{3},0}\;\}
\nonumber
\end{eqnarray}

where $a = (a_{0},a_{1},a_{2},a_{3}) = (\pm1,0,0,0),(\pm1,\pm1,0,0),
(\pm1,0,\pm1,0),(\pm1,0,0,\pm1)$,
$(\pm1,\pm1,\pm1,0),(\pm1,\pm1,0,\pm1),(\pm1,0,\pm1,\pm1),(\pm1,\pm1,\pm1,\pm1)$.

\begin{figure}
\caption{The lattice description used throughout this work. \label{fig1}}
\end{figure}

\begin{figure}
\caption{The behavior of
$P(x,t)$ for $t = 100$ and $-50 \leq x \leq 50$ for closed by
rational and irrational values of $\alpha$:
{\bf a)} $\alpha = 1/3$ (full line) and
          $\alpha = 1/8^{1/2}$ (broken line),
{\bf b)} $\alpha = 0.6$ (full line) and $\alpha = (5^{1/2} - 1)/2$ the
          golden mean (broken line). \label{fig2}}
\end{figure}

\begin{figure}
\caption{ The behavior of
$P(x,t)$ for $x=0$ and $0 \leq t \leq 100$ for $\alpha = 0.6$
and $(5^{1/2} - 1)/2$.\label{fig3}}
\end{figure}

\begin{figure}
\caption{ Comparison of $P(x,t)$ for $t=100$ and $-50 \leq x \leq 50$ for the
pure case (full line) and the random case (broken line). The disorder
average is done over $500$ realizations.\label{fig4}}
\end{figure}

\begin{figure}
\caption{ Log-log plots of $[\langle x^{2} \rangle]$ and
$[\langle x \rangle^{2}]$ for $t \leq 500$.
The former fits a linear $t$ behavior within $10^{-3}$ accuracy. The expected
$t^{\frac{1}{2}}$ asymptotic behavior of the latter is
also depicted (full line).\label{fig5}}
\end{figure}

\begin{figure}
\caption{ Some typical configurations contributing to
$W_{2}$, $W_{3}$ and $W_{4}$.
Arrows' directions indicate paths which belong to $A_{i}$ or to $A^{\ast}_{i}$.
Also shown the relations
between  the coordinate set $(x_{1},\ldots\;,x_{n})$ and the set
$(x,r_{1},\ldots\;,r_{n-1})$.\label{fig6}}
\end{figure}

\begin{figure}
\caption{ All possible steps configurations
of noninteracting pairs of paths
advancing from $t-1$ to $t+1$.\label{fig7}}
\end{figure}

\begin{figure}
\caption{ Log-log plot for moments
of the  probability distribution versus $t$,
compared with their conjunctured large $t$ behaviors (full lines).\label{fig8}}
\end{figure}

\begin{figure}
\caption{ Log-log plot for the probability
that $n$ paths will end at the same point $x=0$
after a distance $t$. The full lines indicate the conjunctured
large $t$ behaviors.\label{fig9}}
\end{figure}
\end{document}